%% file: Paper-PR-D-19-01336.tex
\documentclass[review]{elsarticle}

\usepackage{lineno}
\usepackage[colorlinks]{hyperref}
\hypersetup{pdfauthor=author}
\modulolinenumbers[5]

\journal{Pattern Recognition}

\usepackage{booktabs}
\usepackage{makecell}
\usepackage{tabularx}
\usepackage{ctable}
\usepackage{multirow}
\usepackage{array}
\usepackage{xparse}
\usepackage{color}
\usepackage{xcolor}
\usepackage{xspace}
\usepackage[normalem]{ulem}
\usepackage{graphicx}
\usepackage{lipsum}
\usepackage{eurosym}

\usepackage{wrapfig}

\usepackage{amsmath,amsfonts,amssymb}

\newcommand{\ReviewComment}[4]{%
	\if\relax\detokenize{#4}\relax
		{\textcolor{#2}{#3}}%
	\else%
		{\textcolor{#1}{\sout{#3}}}%
		{\textcolor{#2}{#4}}%
	\fi%
}
\definecolor{darkpink}{rgb}{0.91, 0.33, 0.5}
\definecolor{lightolive}{rgb}{0.8, 0.8, 0.25}
\definecolor{limegreen}{rgb}{0.20,0.80,0.50}

\usepackage{graphicx}
\usepackage{caption}
\usepackage{subcaption}

\usepackage{float}

\renewcommand{\vec}[1]{\mathbf{#1}}

\newcommand{\odd}[1]{\mathcal{#1}_{ODD}}
\newcommand{\pdd}[1]{\mathcal{#1}_{PDD}}
\newcommand{\npdd}[1]{\mathcal{#1}_{NPDD}}

\DeclareMathOperator*{\argmin}{arg\,min}

\usepackage{footnote}
\makesavenoteenv{tabular}
\makesavenoteenv{table}

\newcolumntype{L}{>{\centering\arraybackslash}m{3cm}}

\makeatletter
\def\ps@pprintTitle{
	\let\@oddhead\@empty
	\let\@evenhead\@empty
	\def\@oddfoot{\hfill\thepage\hfill}
	\def\@evenfoot{\thepage\hfill}
}
\makeatother

\let\linenumbers\nolinenumbers\nolinenumbers

\usepackage[margin=3cm]{geometry}

\AtBeginDocument{%
\hypersetup{
	linkcolor = blue,
	citecolor = blue,
	urlcolor  = blue
}
}
\begin{document}

\input{sections/00-title-authors-abstract.tex}

\linenumbers

\input{sections/01-introduction.tex}
\input{sections/02-related-works.tex}
\input{sections/03-proposed-method.tex}
\input{sections/04-experimental-methodology.tex}
\input{sections/05-experimental-results.tex}
\input{sections/06-conclusion.tex}

\bibliographystyle{cas-model2-names}
\bibliography{Paper-PR-D-19-01336}

\end{document}

%% file: sections/00-title-authors-abstract.tex
\begin{frontmatter}
	
	\title{Copycat CNN: Are Random Non-Labeled Data Enough to Steal Knowledge from Black-box Models?}

	\author[ufes]{Jacson Rodrigues Correia-Silva\corref{mycorrespondingauthor}}
	\cortext[mycorrespondingauthor]{Corresponding author}
	\ead{jacson.silva@ufes.br}
	
	\author[ufes]{Rodrigo F. Berriel}
	\author[ufes]{Claudine Badue}
	\author[ufes]{Alberto F. {De Souza}}
	\author[ufes]{Thiago Oliveira-Santos}
	
	\address[ufes]{
		Universidade Federal do Esp\'irito Santo (UFES), ES, Brazil\vspace{1cm}
		\\Link to the formal publication: \url{https://doi.org/10.1016/j.patcog.2021.107830}\vspace{0.2cm}\\
		\begin{minipage}[c]{0.2\textwidth}
			\centering{\includegraphics[width=0.85\textwidth]{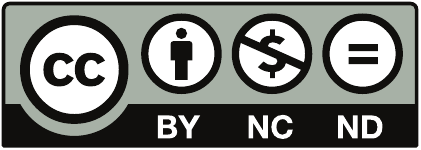}}
		\end{minipage}\hfill
		\begin{minipage}[c]{0.8\textwidth}
			\centering{
				\textcopyright~2021. This manuscript version is made available under the CC-BY-NC-ND 4.0 license\\
				\url{http://creativecommons.org/licenses/by-nc-nd/4.0/}
			}
		\end{minipage}
	}

%	\address[ufes]{Universidade Federal do Esp\'irito Santo (UFES), ES, Brazil}

	\begin{abstract}
		Convolutional neural networks have been successful lately enabling companies to develop neural-based products, which demand an expensive process, involving data acquisition and annotation; and model generation, usually requiring experts.
		With all these costs, companies are concerned about the security of their models against copies and deliver them as black-boxes accessed by APIs.
		Nonetheless, we argue that even black-box models still have some vulnerabilities.
		In a preliminary work, we presented a simple, yet powerful, method to copy black-box models by querying them with natural random images.
		In this work, we consolidate and extend the copycat method: (i) some constraints are waived; (ii) an extensive evaluation with several problems is performed; (iii) models are copied between different architectures; and, (iv) a deeper analysis is performed by looking at the copycat behavior.
		Results show that natural random images are effective to generate copycats for several problems.
	\end{abstract}
	
	
\end{frontmatter}

%% file: sections/01-introduction.tex
\section{Introduction}
\label{sec:introduction}
Convolutional Neural Networks (CNNs) have been extremely successful in a wide range of tasks, such as image classification, object detection, and segmentation. Like most Deep Neural Networks (DNNs), one of their major strength is also an important drawback: they are data-intensive.
These models have been able to achieve those impressive results largely because of the increasing amount of available data.
On the other hand, acquiring and annotating all these data has a high cost.
Nonetheless, on top of this success, many companies have been developing CNN-based products.
Trying to protect their models against copies, it is common to make them available via an Application Programming Interface (API) to be consumed by customers such as, for example, a company selling a robot with a programmable interface and with built-in functionalities for visual pattern recognition. By owning a robot, it is possible to query it an unlimited number of times. Sometimes, usually with the more general models, companies make them available on the cloud (e.g., Microsoft Cognitive Services Computer Vision API, Google Cloud Vision API, IBM Watson Visual Recognition, etc.) or provide them as part of a framework or a product to be consumed as a black-box by whoever is willing to pay per usage.

There are at least three high costs associated with a model:
(i) acquiring and annotating large-scale training datasets; (ii) computational power for training the models that can last for days or even months; and (iii) experts to prepare the data and to design, implement, and train the model.
With all these costs, companies usually do not deliver their models as white-boxes (with the parameters, architecture, and other details visible outside the model) but as black-boxes instead, given that otherwise it would allow costumers to just copy the model and stop paying for its usage or using it to generate competing products. To be more restrictive, some companies provide only the final label of the model instead of the classes probabilities since the probabilities could lead to some additional vulnerabilities. For instance, distillation-based~\cite{hinton2015neurips} techniques could be employed to distill the knowledge embedded in the parameters of a model into another (potentially smaller) model. Such an approach, however, would require two things: access to the class probabilities of the black-box and images from the problem domain, i.e., the same domain used to train the model of interest. Some works like ~\cite{tramer2016usenix} have already investigated the possibility of copying models by querying them as black-boxes.
Nonetheless, they did not explore DNNs nor investigate copies without data from the problem domain to query the model of interest which imposes an additional strong restriction to perform the copy.

Pushing the assumptions of~\cite{hinton2015neurips} even further, we hypothesized that one may not even need data from the problem domain to successfully copy a model. In \cite{nguyen2015cvpr}, for instance, the authors showed images completely unrecognizable to humans fooling state-of-the-art DNNs and obtaining results with $99\%$ of confidence.
Along this line, our hypothesis is that a set of natural random images (not necessarily related to the problem domain) can be used to gather the necessary knowledge from a target black-box model in order to generate a similar copycat model. Furthermore, we also hypothesize that it is possible to perform the copy using only the hard label outputs (i.e., one-hot) of the model instead of the classes probabilities. In a preliminary work~\cite{jacson2018ijcnn}, our group presented a simple, yet powerful, method to copy a black-box CNN model by querying it with data out of the problem domain, i.e., using natural random images. 
Preliminary experiments showed promising results with the copycat model achieving accuracies close to the target black-box model. The main advantage of the copycat is that it works with natural random images (not related to the problem domain) and hard labels instead of classes probabilities.

This work consolidates the copycat method by performing an extensive evaluation study aiming at providing a better understanding of the behavior of the network on black-box attacks with natural random images and hard labels. Therefore, a set of experiments are performed in order to try to explain the behavior and limitations of such cloned networks by using methods of network analysis, like Layer-wise Relevance Propagation (LRP) \cite{bach2015plos}. Such a study for explaining the vulnerabilities of DNN models enables the pattern recognition community to better plan defenses against attacks. This work has many differences to the preliminary work that can be highlighted as follows. (i) To corroborate with our hypothesis that natural random images are sufficient to generate copycat models, this work starts performing a new qualitative analysis of the two distributions (natural random images out of the problem domain and images from a specific problem domain) in the classification space of a given target black-box model. (ii) This work waves some constraints and shows that copy attacks can be performed with even less information about the black-box model. Differently from the preliminary work, the subtraction of training images' mean (assumed to be known) are not used on copycat models, and the fake datasets are simply balanced by random replication of their own images.
(iii) The copy attack is extensively evaluated with more problems of diverse domains. An analysis is performed to check the influence of the number of classes and the need for natural images against images produced with random pixels. (iv) An analysis is performed on the relation between the amount of natural random images used to copy the model and the achieved model accuracy. (v) In order to see the limitations of copycat attack to smaller architectures, evaluation is performed with the method used to transfer knowledge from models with different architectures. (vi) The robustness of the copycat model is also analyzed against the random factors of the black-box model training. (vii) The Layer-wise Relevance Propagation (LRP) \cite{bach2015plos} is used to better interpret the model by showing a visual comparison of the image regions that the copycat and the target black-box models are considering for predictions. (viii) Finally, the conditions for the viability of a realistic attack are discussed and the costs of such attack to a real-world API is analysed. Results show that natural random images labeled only by hard labels are effective to generate copycat models with functionalities similar to those of the target model for a variety of problems.

%% file: sections/02-related-works.tex
\section{Related Works}
Since the success of deep neural networks, there has been an increasing concern about the security of the models. Therefore, several works have been exploring methods to attack state-of-the-art models. These attacks on the deep networks usually aim at coping the model knowledge (knowledge distillation and extraction attacks) or at causing classification errors (adversarial examples). Since these attacks are closely related to this work, this section presents an overview of the Knowledge Distillation, Adversarial Examples, and the Vulnerabilities and Attacks of the models.

\subsection{Knowledge Distillation}
Research on knowledge distillation shows that knowledge can be obtained from the model logits or classes probabilities. In fact, as discussed in \cite{hinton2015neurips}, not only the output label, but also the logits provide information about the model, therefore many works have tried to explore it. 

\citet{bucila2006kdd} used logits as pseudo-labels to train a new neural network.
At that time, they used images from the problem domain to generate new synthetic ones and argued that it is complicated to find large amounts of data for some domains. They also mentioned that it is necessary that synthetic data matches as well as possible the distribution of real train and test data, covering most of the region of interest in the input space.
For this purpose, they proposed a method that generates synthetic data by selecting pairs of closest examples and interpolating their own attributes by using a probability parameter.
This method is able to generate samples closer to the real data distribution.
Likewise,~\citet{ba2014neurips} showed that deeper and complex convolutional architectures could be \textit{compressed} in shallow fully-connected networks, but they still rely on the availability of the logits.

Later, the distillation process was formalized by \citet{hinton2015neurips} by defining a softmax with temperature to obtain softer probability distribution over classes.
The proposed method was used to transfer knowledge from an ensemble of models and from a large model to a smaller one by using their logits tuned by the temperature. The authors show that it is easier to extract knowledge out of the model by using soft labels instead of hard labels. However, this imposes additional constraints if the method is used for copy attacks. Some other works were built on the top of these results. For example, \citet{tang2016icassp} employed the distillation to extract knowledge from a simpler DNN to train an LSTM.

Unlike our work, however, these methods usually assume that some details about the models are known, i.e., they treat the model of interest as a white-box.
In some of these works, the architecture is assumed to be known; in other works they require to use the same training data (or problem domain data, at least) used by the model of interest; and in most of them it is assumed that one would have access to the logits or probabilities of all classes for a given input.

\subsection{Adversarial Examples}
The knowledge and predictions of the target models are also explored by adversarial examples.
Many works use the soft labels or the hard labels to craft attacks that aim to cause the misclassification of the oracle. Most of these works apply some noise on input image
capable of changing the oracle output to a chosen output.
Among several works of adversarial examples, this section summarizes the more closely related to our research.

\citet{szegedy2014iclr} showed that machine learning models, including MLP networks, have blind spots and are vulnerable to misclassification attacks. They modified an already correctly classified image by adding noise that is imperceptible to human eyes. Nonetheless, the almost identical image was confidently misclassified by the models. On the other hand, \citet{nguyen2015cvpr} showed that although deep networks have achieved competitive results compared to humans, they perceive objects in a different manner. For example, the networks provide classifications with high level of confidence even to images that are unrecognizable to humans. They found that genetic algorithms are able to find images that are far from the training set space, but with features capable of misleading the classification of a model, i.e., they produce highly-confident results for images that are far from the problem domain space used in the training set. With such images the model could be fooled to classify images in a specific way.
Moreover, researchers also showed that these models can be fooled by a very small amount of noise. 

\citet{goodfellow2015iclr} proposed using adversarial examples to expose blind spots caused by training algorithms. By fast gradient sign method, they caused the misclassification in several models. Even models trained with different architectures, but on same dataset, were vulnerable to the same adversarial example.
It also was observed that adversarial examples are a result of models being too linear, rather than too nonlinear, and of spaces not covered on the training. In addition, they point out that easy-to-optimize models are easier to attack.

In response to these initiatives, there is a growing interest in building defense against these adversarial examples. The reader is referred to \cite{yuan2019tnnls} for more recent works on adversarial attacks and defense.

\subsection{Vulnerabilities and Attacks}
With respect to our attack of interest, i.e., copy attack of a target black-box model, there are some works exploring similar approaches. In~\cite{shi2017hst}, the authors frame these as \textit{exploratory} attacks and set out to demonstrate that deep learning models can steal the functionalities of two classifiers (Naive Bayes and SVM). However, their work did not explore copies from DNNs, besides demanding problem domain data. By investigating extraction attacks, \citet{tramer2016usenix} explored vulnerabilities in machine learning APIs. They aimed at copying the functionality of black-box models (e.g., logistic regression, SVM, decision trees, and MLP) just by querying the models of interest that were available on cloud services such as BigML and Amazon. Although investigating copy attacks using only the target model hard labels, their work focused on using images from the problem domain and did not investigated copies from DNNs.

~\citet{papernot2017asiaccs} used a black-box target DNN to label a small set of the MNIST. Subsequently, they trained a set of substitute DNN models using this dataset. The Jacobian matrix was applied on each model to generate new samples aiming at finding the boundaries of the target DNN. Therefore, the substitute models were used as white-box to generate the synthetic images. After labeling the new images using the target DNN, the models were trained again until achieving an accuracy sufficient to craft adversarial examples. A refinement to this technique was proposed in~\cite{papernot2016arxiv}. 
Finally, the authors transferred the knowledge from machine learning classifiers (DNN, logistic regression, SVM, decision tree, nearest neighbor, and ensembles) to a deep model. Their experiments were limited to the MNIST dataset though.

More recently, and after our preliminary work, \citet{orekondy2019cvpr} studied how an adversary can steal functionalities of a black-box target network. Similarly to our preliminary work~\cite{jacson2018ijcnn}, the authors labeled a set of natural random images using a black-box model to generate a fake dataset. But unlike our work, they used the probabilities (soft labels) instead of hard labels to label their fake dataset.
They conducted experiments with four datasets and a fixed architecture (ResNet-34) with pre-trained weights. The experiments followed three data distributions: images used to train the target networks but unlabeled; a combination of all images used to train all target networks, images from OpenImages \cite{kuznetsova2018arxiv}, and images from ILSVRC \cite{russakovsky2015ijcv}; and images only from OpenImages and ILSVRC.
Furthermore, they applied one attack method with random image samples and another one with reinforcement learning. Additionally, to verify the influence of the architecture, they generated target networks with VGG-16 and ResNet-34 architectures trained with one of the datasets. Subsequently, they used several architectures to attack the target networks.
Finally, they concluded that it is beneficial to use a more complex architecture to copy the network and achieved more than 77\% of copy in their experiments.

Although similar, their work have some key differences to ours. In \cite{papernot2017asiaccs}, a substitute model was used to create adversarial examples, but the model requires images from the problem domain images. Additionally, the authors aimed at copying the boundaries of the target network in order to create adversarial examples and not copying the model itself. In \cite{orekondy2019cvpr}, the attack assumes access to the probabilities (soft labels) and not just the hard label output.

Along the line of this work, \citet{mosafi2019icann} presented an attack with unlabeled data over models trained with MNIST and CIFAR10. The authors used the probabilities of target network to label the unlabeled data. In a follow up work \cite{mosafi2019ijcnn}, they cite our preliminary work and decide to use the same constraints, i.e., only the hard labels. Additionally, they proposed a new method to create unlabeled data, but a deep analysis of the model behavior was not performed.

To the best of our knowledge, our preliminary work was the first to steal knowledge from the target model using only natural random images and final model labels. The current work explores this type of copy even further by extending the experimentation and by performing a novel model behavior analysis. Unlike related works that use final model probabilities, the proposed method requires only access to the hard labels and is tested with different problem domains comprising different number of classes and in different architectures. These experiments allows to present an initial analysis of the limitations and capabilities of copying with natural random images and final labels of the model. The \autoref{tabRelatedWorksComparison} show a comparison between related works and ours (Copycat).

\begin{table}[h]
	\caption{Comparison between related works and Copycat (ours). Abbreviations: Distillation (D), Adversarial Examples (A) or Copy Attack (C), Problem Domain Data (P), and Non Problem Domain Data (N).
	}
	\label{tabRelatedWorksComparison}
	\resizebox{\textwidth}{!}{
		\begin{tabularx}{1.3\textwidth}
			{>{\hsize=.35\hsize}X|
				>{\centering\hsize=.03\hsize}X|
				>{\centering\hsize=.03\hsize}X|
				>{\centering\hsize=.03\hsize}X|
				>{\centering\hsize=.03\hsize}X|
				>{\centering\hsize=.03\hsize}X|
				>{\centering\hsize=.03\hsize}X|
				>{\centering\hsize=.03\hsize}X|
				>{\centering\hsize=.03\hsize}X|
				>{\centering\hsize=.05\hsize}X|
				>{\centering\hsize=.03\hsize}X|
				>{\centering\hsize=.05\hsize}X}
			\toprule
			Features~/~Methods &
			\cite{bucila2006kdd} &
			\cite{hinton2015neurips} &
			\cite{tang2016icassp} &
			\cite{szegedy2014iclr} &
			\cite{goodfellow2015iclr} &
			\cite{nguyen2015cvpr} &
			\cite{shi2017hst} &
			\cite{tramer2016usenix} &
			\cite{orekondy2019cvpr} &
			\cite{mosafi2019ijcnn} &
			Ours
			\tabularnewline \midrule
			Type of method          & D & D & D & A & A & A &   & C & C & C & C  \tabularnewline
			\midrule
			Black-Box               & \checkmark & \checkmark &   &   &   & \checkmark & \checkmark & \checkmark & \checkmark & \checkmark & \checkmark \tabularnewline
			\midrule
			Data used in the Experiments & P & P & P & P & P & N & P & P & N,P & N & N,P \tabularnewline
			\midrule
			Hard labels             &   &   &   &   &   &   & \checkmark & \checkmark &   & \checkmark & \checkmark \tabularnewline
			\midrule
			ML, Shallow Networks    & \checkmark &   &   & \checkmark & \checkmark &   &   & \checkmark &   &   &   \tabularnewline
			\midrule
			DNN $\rightarrow$ DNN   &   & \checkmark & \checkmark & \checkmark & \checkmark & \checkmark &   &   & \checkmark & \checkmark & \checkmark \tabularnewline
			\midrule
			ML $\rightarrow$ DNN    &   &   &   &   &   &   & \checkmark & \checkmark &   &   &   \tabularnewline
			\bottomrule
		\end{tabularx}
	}
\end{table}

%% file: sections/03-proposed-method.tex
\section{Copycat Convolutional Neural Network}
\begin{figure}[h]
	\centering
	\includegraphics[width=\linewidth]{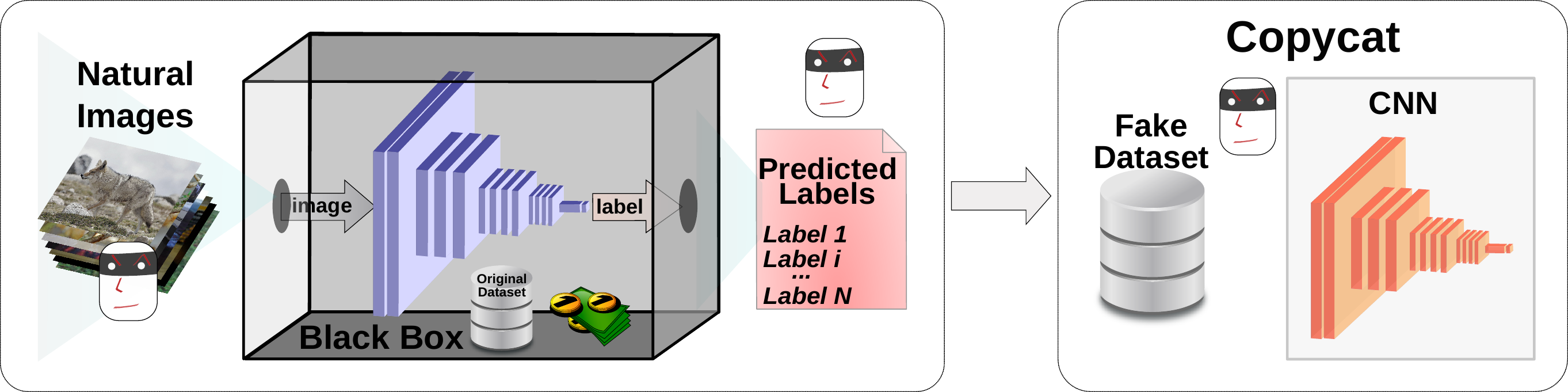}
	\caption{
		Overview of the copycat creation. The target black-box model is created using proprietary data (original dataset) from companies that are not willing to share the model for free. The model is trained to work on a specific problem domain. The attacker wants to generate a copycat model without paying. Therefore, the attacker queries the black-box model with natural random images available on the internet in order to obtain the predicted hard labels (within the original problem domain) for the natural images. The pairs of natural images and predicted labels comprise the fake dataset which is used to train a copycat model of the black-box. The copycat is expected to have similar accuracy to the black-box model.
	}
	\label{fig:method}
\end{figure}

\autoref{fig:method} illustrates the overview of our method to copy a target black-box model using natural random images. In summary, the attacker gathers a dataset of natural random images, queries the target black-box model to generate their labels in the black-box model classification space, and, finally, uses those pairs (natural image and predicted label) as a fake dataset to train a copycat model.
 
The remainder of this section first walks through the formulation of different types of attack and later presents the general details of the copycat training.

\subsection{Attack formulation}
This subsection starts from the definition of the model of interest and goes until the formulation of the copy attack with natural random images by adding constraints that are reasonable in this type of attack. 

Let $f$ be a deep neural network model parametrized by $\theta$, then:
\begin{equation}
\vec{y}_i' = f_{\theta}(\vec{x}_i),
\end{equation}
where $\vec{x}_i$ is an input sample and $\vec{y}_i' \in \mathbb{R}^K$ is the soft-labels (predictions) of the model after a softmax, i.e., the vector of probabilities for each class $k \in K$. Let $\odd{X} = \{\vec{x}_i\}_{i=1}^{N}$ 
be the $N$ image samples and 
$\odd{Y} = \{\vec{y}_i\}_{i=1}^{N}$ their corresponding one-hot labels (representing the class of $\vec{x}_i$), which together comprise the \textit{original domain dataset (ODD)} used to train $f_{\theta^{*}}$. The trained model of interest, $f_{\theta^{*}}$, can be obtained by:
\begin{equation}
f_{\theta^{*}} = \argmin_\theta \mathcal{L}(f_\theta(\vec{x}_i),\vec{y}_i).
\end{equation}
where $\mathcal{L}$ is a multi-class classification loss such as Cross-Entropy.

Often, $\odd{X}$ and $\odd{Y}$ are private data that no company is willing to share. Therefore, one can assume that the attacker does not have access to the ODD. This already differs from many of the distillation-based methods. Nonetheless, it is fair, in this case, to assume that the attacker may still know the domain in which the model operates (for example, one may know that the model of interest predicts breeds of dogs). With this knowledge, it is also fair to assume that an attacker may be able to get images from the domain of interest (referred here as \textit{problem domain dataset -- PDD}). The acquisition and the annotation of such data is costly. Assuming that the attacker can afford such costs and that the attacker knows the architecture of the model, then one could try to train $f$ using this newly acquired PDD, such that:
\begin{equation}
	f_{\theta^{*}} = \argmin_\theta \mathcal{L}(f_\theta(\vec{\tilde{x}}_i), \vec{\tilde{y}}_i),
\end{equation}
where $\pdd{X} = \{\vec{\tilde{x}}_i\}_{i=1}^{M}$ and $\pdd{Y} = \{\vec{\tilde{y}}_i\}_{i=1}^{M}$. As the attacker has costs to acquire and annotate this data, it is also reasonable to assume $M \ll N$.

As these models are really expensive, the companies try to protect them at all costs. Therefore, it is also likely that the attacker does not know the architecture of the model to be copied and would have to copy the model to an architecture that is probably different from the black-box. The training of a model with a different architecture could be represented as:
\begin{equation}
	g_{\phi^{*}} = \argmin_\phi \mathcal{L}(g_\phi(\vec{\tilde{x}}_i), \vec{\tilde{y}}_i),
\end{equation}
where $g_{\phi^{*}}$ is a different model ($g$) with different set of parameters ($\phi^{*}$) and the complexity of the model can be chosen according to the constraints of the attacker (e.g., computational resources, time, performance requirements, etc.). These cases, however, cannot be considered attacks but just competing models. 

The formulations so far are equivalent to cross-databases experiments performed in the literature and they have no relation to the model of interest. In a more realistic scenario, the attacker wants to specifically mimic the model of interest while reducing the annotation costs of the attack, i.e., the attacker does not want to annotate the set of images to train the model. In this case, the attacker may only acquire samples ($\pdd{X}$) from the problem domain but not their corresponding labels ($\pdd{Y}$); for instance, by crawling images on the web. Removing the cost of annotating the data is an important step to make such attacks feasible, given that this process is sometimes more expensive than the acquisition of the data itself; even more expensive than the access to the model of interest in some cases. To mimic the behavior of the model of interest, the following formulation can be used:
\begin{equation}
\label{eq:formulation-pdd}
	g_{\phi^{*}} = \argmin_\phi \mathcal{L}(g_\phi(\vec{\tilde{x}}_i), f_{\theta^{*}}(\tilde{\vec{x}}_i)).
\end{equation}
Note that the prediction of the model of interest $f_{\theta^{*}}(\tilde{\vec{x}}_i) \in \mathbb{R}^K$ is a vector of probabilities (soft-labels). The formulation in \autoref{eq:formulation-pdd} can be considered a copy attack with access to the soft-labels, because $g_{\phi}$ will be trained to mimic the black-box model $f_{\theta^{*}}$ by minimizing the difference between both predictions.

This work, however, pushes these constraints even further. We hypothesize that, in order to train $g_{\phi}$ to steal the functionalities of the model of interest $f_{\theta^{*}}$, (i) there is no need for data from the problem domain and (ii) there is no need to have access to the soft-labels, i.e., the one-hot (hard-labels) are sufficient. Therefore, the hypothesis of this work can be modeled by the formulation:
\begin{align}
\label{eq:formulation-npdd}
g_{\phi^{*}} &= \argmin_\phi \mathcal{L}(g_{\phi}(\hat{\vec{x}_i}), \sigma(f_{\theta^{*}}(\hat{\vec{x}}_i)))\\
\sigma(\hat{\vec{y}}) &= 
\begin{cases}
	1& \hat{y_i} = \text{max}\{\hat{y_i} :\hat{y_i} \in \hat{\vec{y}}\}\\
	0& \text{otherwise}
\end{cases}
\end{align}
where $\hat{\vec{x}}_i$ are image samples that do not belong to the problem domain (referred here as \textit{non problem domain dataset -- NPDD}), with $\npdd{X} = \{\hat{\vec{x}}_i\}_{i=1}^{Z}$,
and $\sigma(\cdot)$ is the one-hot (hard-label) encoding function of the prediction of the black-box model, which provides $1$ for the class with the highest probability and $0$ otherwise.
The images from $\npdd{X}$ are natural random images from classes that are not related to the problem domain. Because of the domain shift and the easiness of acquiring samples, it is reasonable to assume that $Z \gg N$.

Such approach could potentially reduce the difficulty of performing the attack, since the attacker now only needs to acquire natural image samples not necessarily related to the problem domain. This type of image can be easily acquired over the internet and, once collected, it can be used in multiple attacks. 

\subsection{Copycat Model Training}
The training of the model described in \autoref{eq:formulation-npdd} comprises two parts: querying the target black-box model that outputs $\sigma(f_{\theta^{*}}(\hat{\vec{x}}_i))$ and training the copycat network $g_\phi$. When querying the target model with images that are not from the problem domain, several labels for different classes are obtained, but one cannot guarantee that the amount of image samples per class of the problem domain is balanced. It might be the case that for a problem with $\hat{\vec{y_i}} \in \mathbb{R}^K$, many natural images $\hat{\vec{x}}_i$ are predicted with the same label. Differently from \cite{jacson2018ijcnn}, in this work, after obtaining the labels from the black-box model, images are balanced so that the natural image samples are uniformly distributed among the classes of interest. The data balance is achieved by randomly replicating or eliminating images from the classes respectively missing or exceeding the desired number of images. Once the data is balanced, the model is trained. 

It is also common to subtract the mean image (computed on the training dataset) from the inputs during training. This is a common pre-processing step that is also applied during inference; therefore, this image can be considered part of the model. Companies providing an API on the cloud, however, would probably encapsulate such mean image to avoid that the user has access to it. In other words, such image would be part of the black-box model. In this work, differently from \cite{jacson2018ijcnn}, the mean image is considered to be unknown. Therefore, the copycat model is trained without the mean image even when the target model is subtracting the mean image from the queries (which one cannot know from a black-box).

%% file: sections/04-experimental-methodology.tex
\section{Experimental Methodology}
This section describes the materials and methods used for the experimentation of the models.

\subsection{Datasets Organization and Baselines Setup}
Several datasets were used to evaluate the proposed method. First, the datasets used for the baselines alongside the one used to test all the models are presented. Finally, the datasets used for the copycat networks are described.

\subsubsection{Datasets for Baselines and Tests}
\label{secExpMet:subDatBasTes}
To evaluate the proposed approach, three dataset splits were generated for each problem: original domain dataset (ODD), problem domain dataset (PDD) and test domain dataset (TDD). This split was performed for each one of the investigated problems taking into account datasets available in the literature. Specific details about each split of each problem are presented in Section~\ref{subsec:problems}.

The first dataset split is the original domain dataset (ODD) that is used to train the target network. It represents the set of confidential data that is used to train the black-box model of interest. Note that no copycat network has access to this dataset.
The second dataset split is the problem domain dataset (PDD), and it represents the dataset that attackers may be able to build on their own. This dataset comes from the same domain of the ODD but might have less images.
As the PDD tends to be small, an augmentation process can be used to increase the number of samples and to provide a larger and balanced dataset for model training. This augmentation reflects what attackers would do to increase the number of images they have. It is important to note that PDD and ODD do not share a single sample.
Finally, the third dataset is the test domain dataset (TDD) that is used for testing the performance of all models. It comprises images from the problem domain and does not overlap with the ODD and the PDD. None of the networks have access to samples of this dataset.

Given that we performed experiments on the PDD using both the original labels (OL) and the stolen labels (SL), i.e., the labels predicted by the target black-box model, we appended -SL and -OL on each dataset split accordingly. Specifically, all the experiments with the ODD and TDD only used the OL; therefore, they will be always referenced as ODD-OL and TDD-OL, respectively. On the other hand, depending on the experiment, the PDD-OL or the PDD-SL can be used depending on whether the original or stolen labels were used.

\subsubsection{Datasets for Generating the Copycats}
The dataset used to train the copycats comprises images that are not from the problem domain (natural random images) labeled by the black-box model, hence NPDD-SL.
Given that the acquisition of these images is assumed to be inexpensive and that only the hard-labels predicted by the black-box model are used, it is reasonable to assume that, to be effective, this dataset has a lot more samples than the ODD-OL.

To accomplish that, we exploited what is at the disposal of any other attacker: publicly and readily available datasets. 
Based on the need for natural random images, two of the most well-known public datasets of images were used: the ImageNet (Visual Domain Decatlhon\cite{rebuffi2017neurips} and the ILSVRC2017~\cite{russakovsky2015ijcv}) and Microsoft COCO~\cite{lin2014eccv}.
All datasets are merged and have their duplicates removed, leaving 3,088,392 images left. All the original labels are discarded, since this dataset is always used with stolen labels. The number of images per class varies per problem, because it depends on the predictions of the target black-box model. After generating all image-label pairs, the dataset is balanced by oversampling the minority classes and under-sampling the majority classes.

\subsection{Investigated Problems}
\label{subsec:problems}
In order to evaluate the proposed method with different real-world problems, seven problems were chosen:
Human Action Classification (ACT101),
Handwritten Digit Classification (DIG10),
Facial Expression Recognition (FER7),
General Object Detection (GOC9),
Pedestrian Detection (PED2),
Street View House Number Classification (SHN10), and
Traffic Sign Classification (SIG30).
An overview of public domain data used to generate the datasets and the image distribution between them are shown in \autoref{problemDetails}.

\begin{table}[h!]
	\caption{
		Details of the problems, their respective datasets, and domain splits.
	}
	\label{problemDetails}
	\resizebox{1.\textwidth}{!}{%
	\begin{tabular}{lLccc}
		\toprule
		Problems & Datasets & \# ODD & \# TDD & \# PDD \\
		\midrule
		ACT101 & \multicolumn{1}{m{8cm}}{
				 UCF101 \cite{soomro2012crcv}
			     }
  			   & 1782858 & 685824 & 101104 \\
		DIG10 & \multicolumn{1}{m{8cm}}{
				MNIST,
				NIST Special Database 19
				\footnote{MNIST: \url{http://yann.lecun.com/exdb/mnist/}, NIST: \url{http://doi.org/10.18434/T4H01C}}
				}
			  & 60000 & 10000 & 55000 \\
		FER7 & \multicolumn{1}{m{8cm}}{
			   AR Face \cite{martinez1998ar},
		       BU3DFE \cite{yin2006fgr},
		       JAFFE \cite{lyons1998fg},
		       MMI \cite{pantic2005icme},
		       RaFD \cite{langner2010ce},
			   KDEF \cite{lundqvist1998kdef},
			   CK+ \cite{lucey2010cvprw}
			   }
			 & 55629 & 1236 & 65660 \\
		GOC9 & \multicolumn{1}{m{8cm}}{
			   CIFAR-10 \cite{krizhevsky09thesis},
			   STL-10 \cite{coates2011aistats}
			   }
		     & 45000 & 9000 & 269100 \\
		PED2 & \multicolumn{1}{m{8cm}}{
				DamPed \cite{munder2006tpami}
			   }
		     & 23520 & 5880 & 19600 \\
		SHN10 & \multicolumn{1}{m{8cm}}{
				SVHN \cite{netzer2011neurips}
			   }
		      & 47217 & 26040 & 26032 \\
		SIG30 & \multicolumn{1}{m{8cm}}{
			    TT100K \cite{lu2018cvm},
			    TSRD \cite{huang2014pr}
			   }
		      & 31775 & 5481 & 30000 \\
		\bottomrule
	\end{tabular}%
	}
\end{table}

\subsection{Investigated Architectures}
The proposed approach was evaluated in two conditions regarding the network architecture: generating copycats from the same architecture or from different architectures. Two well-known architectures were chosen for this experiments: the VGG-16 \cite{simonyan2015iclr} and the AlexNet \cite{krizhevsky2012neurips}.
The VGG-16 was chosen because the model achieves high accuracies in all problems investigated, whereas the AlexNet is a well-known model with limited learning capacity.

\subsection{Metric}
To evaluate the proposed method, quantitative and qualitative metrics were used.
To measure the performance of the network, the accuracy was calculated by its macro average on the test domain dataset.
Since data imbalance may exist in a dataset, this metric is useful to handle this peculiarity.
It computes the simple average over the classes independently and then calculates the average, hence treating all classes equally.
Furthermore, the ratio between the copycat accuracy over target network and over the alternative baseline are also used to measure the network performance.
The perfect copycat network should achieve a macro-averaged accuracy identical to the target network (i.e., it should hit and miss the same images as the target network) and achieve $100\%$ of copy performance over the target network.

The Layer-wise Relevance Propagation (LRP) \cite{bach2015plos} was chosen for a qualitative analysis.
It explains individual neural network predictions in terms of input variables.
The LRP is a method for identifying important pixels that contributed most strongly to the image classification.
The method runs a backward pass, and generates per-pixel scores based on their relevance in classification.
In the end, it generates a heatmap with the scores of the pixels.
By using LRP, both target network and copycat heatmaps can be obtained from the same input image.
After, the heatmaps of both networks can be compared to verify their similarity.
We believe that, in general, the same pixels responsible for the prediction of the target network will be highlighted by the copycat network.

\subsection{Experiments}
A set of six experiments are performed to evaluate the proposed method: Analysis of the Input-Images Space in the Black-Box Model, Copycat from the Same Architecture, Curve of Different Sizes of Training Datasets, Copycat from a Different Architecture, Robustness of the Copycat Model, and Analysis of the Attention-Region in the Input Images, and Analysis of Attack Viability and APIs Costs. Each of these experiments is explained in details below. 

\subsubsection{Analysis of the Input-Images Space in the Black-Box Model}
The proposed method aims to copy the target network using the knowledge extracted from its labels.
However, the copycat's performance is directly related to the details provided by the target network about its internal knowledge, i.e., its classification space that was learned during the training.
Therefore, this experiment attempts to visually analyze whether the NPDD-SL images belong to the same space as the ODD-OL images when processed by the target network.

In this direction, we extract the vector generated by the ReLU layer before the logits in the VGG architecture. 
Due to the high dimensionality of 4096, we use T-SNE to generate a 2-D view.
To reduce complexity when running T-SNE, we randomly selected a few examples to perform this experiment: 100 for each class of the ODD-OL.
For each ODD-OL sample, three neighbors of the same class of NPDD-SL were selected.
When a neighbor has already been selected by another point, it is ignored and the next one is selected.
For better visualization and based on the results discussed further in \autoref{secDataCurve}, the neighbors were selected from a random subset of 200k samples of NPDD-SL. Furthermore, before applying T-SNE, the ODD-OL and NPDD-SL vectors were standardized using, respectively, the ODD-OL and the NPDD-SL means and standard deviations.

The analysis is performed for two problems, and the expectation is that several NPDD-SL samples will be in the same space that ODD-OL samples. If this is the case, we consider that they are in the same space inside the model.

\subsubsection{Copycat from the Same Architecture}
In this experiment, the purpose is to copy the target network using the same architecture (VGG).
Initially, the target network is trained with ODD-OL and the alternative baseline is trained with PDD-OL.
Then, the target network is used as a black-box and only receives an image as input, providing a hard-label as output.
At this point, all NPDD and PDD images are queried to the target network and their labels are stolen to create the fake datasets NPDD-SL and PDD-SL.
Next, one copycat is trained with NPDD-SL and another with PDD-SL. Additionally, the first one is fine-tuned with PDD-SL, generating the copycat NPDD+PDD-SL.
Finally, all networks are evaluated using the TDD-OL and their accuracies are compared.
This process is replicated for all problems.

The alternative baseline and the copycats receive an acronym in the format: [BL or CC]-[Architecture]-[Dataset]-[OL or SL].
Therefore, the baselines are: BL-VGG-ODD-OL (target model) and BL-VGG-PDD-OL.
And, the copycats are: CC-VGG-NPDD-SL, CC-VGG-PDD-SL, CC-VGG-NPDD+PDD-SL.

\subsubsection{Curve of Different Sizes of Training Datasets}
This experiment analyzes the differences among copies of the network with different amounts of natural random images. Therefore, datasets with different sizes are created with random images from NPDD-SL. For each problem, copycat networks are trained with NPDD-SL comprising 100k, 500k, 1M, 1.5M, and 3M images. After training, these networks are fine-tuned with PDD-SL.
Finally, all the two versions of the networks (with and without fine-tuning) are evaluated with TDD-OL and their accuracies are used to create a data curve.

\subsubsection{Copycat from a Different Architecture}
To verify the possibility to copy the network knowledge to another architecture, a different one is used to evaluate the proposed method.
For this, a smaller but classic architecture is chosen: AlexNet.
In this experiment, the copycat network is trained with the same NPDD-SL used by CC-VGG-NPDD-SL, i.e., the same pairs of image and stolen label obtained by the target network.
Since the copycat may achieve a lower accuracy due to the architecture capacity and not because of the copy method itself, an alternate baseline (BL-Alex-ODD-OL) is also trained with the same data (ODD-OL) used in the target network.

\subsubsection{Robustness of the Copycat Model}
A neural network initiates the training with random parameters and the samples are also randomly selected, so two or more models trained with the same architecture and the same dataset can achieve different results depending on the initialization procedure.
Since the performance may be different for each training, the copycat method was evaluated on three different target networks trained with the same problem domain images. For each one, the weights were initialized with different values and the images were shuffled on training.
Then, the proposed method is applied to copy each model and the robustness of the copycat against the random factors of the black-box model training is analyzed.

This experiment is only performed on the FER7 using the VGG architecture.
Then, three target networks are generated using the ODD-OL with one copycat being trained (using 3M of NPDD-SL) for each of them. Finally, all networks are evaluated using the TDD-OL and analyzed by comparing their performance variation. It is expected that the three copycats have similar performance. The average and the standard deviation of this performance are used as metric.

\subsubsection{Analysis of the Attention-Region in the Input Images}
To better understand the models and also to analyze the results of the proposed method, a visual comparison of the input image regions that influence the model predictions was performed between copycats and target networks.
For this process, the Layer Relevance Propagation was used\footnote{Due to differences between the LRP toolbox Caffe and NVIDIA Caffe, only images that provided the same predictions in both frameworks were used.}.
After applying LRP on different images, the heatmaps of the following predictions of the target network and copycat are compared: correct vs. correct, correct vs. wrong, and wrong vs. correct.
To analyze the differences between the copycat on different amounts of data, the heatmaps are also generated to copycats used in the data curve (\autoref{secDataCurve}). An image with the same prediction by the target network and copycat is expected to present similar heatmaps for both networks.

\subsubsection{Analysis of Attack Viability and APIs Costs}
In real-world, consuming an API has a per-use cost, which implies that executing an attack as the one formulated in our work will cost money for the attacker. Moreover, for the attack to be successful, one would only consider copies with a certain level of quality. Therefore, we analyse the viability of a realistic attack by measuring the costs associated with different number of queries and their corresponding copy performances. For this experiment, the Emotion API from Microsoft Azure Cognitive Services\footnote{\url{https://azure.microsoft.com/en-us/pricing/details/cognitive-services/emotion-api/}} was used and the performances were measured by the Macro Average on the Test Domain Dataset (TDD). Both TDD and the number of expressions are the same as those used in FER7.

Given that an attack might be viable, we also analyse how much an API should charge to inhibit this type of attack. From the attacker's perspective, an attack is viable when the costs of executing it are lower than acquiring and annotating data of the problem of interest. In this analysis, the price for annotating the data was based on the Data Labeling Service from Google Cloud\footnote{\url{https://cloud.google.com/ai-platform/data-labeling/pricing}}, the amount of data was based on the ODD of each problem, and we consider an attack to be successful with a copy performance greater than $90\%$.

\subsection{General Setup}
\label{sec:setup}
All models were trained using Stochastic Gradient Descent (SGD) with a Step Down policy for the learning rate.
The maximum of five epochs was chosen because the models empirically showed convergence after it.
However, the training of the target networks lasted for more than five epochs, until their losses reached a plateau.
The target network was trained with the corresponding original domain dataset (ODD-OL) for 30 epochs in FER, 20 epochs in GOC, and 5 epochs in other problems.
In the augmentation process, the following operations were applied: add/sub intensity, contrast normalization, crop, horizontal flip, Gaussian blur, Gaussian noise, piecewise affine, rotate, scale, sharpen, shear, and translate.
Code and experiment details are publicly released\footnote{\url{https://github.com/jeiks/Stealing\_DL\_Models}}.

%% file: sections/05-experimental-results.tex
\section{Experimental Results}
This section presents the results of the experiments performed. 
 
\subsection{Analysis of the Input-Images Space in the Black-Box Model}
\begin{figure}[!ht]
	\centering
	\begin{subfigure}{.57\textwidth}
%		\centering
		\includegraphics[width=.98\textwidth]{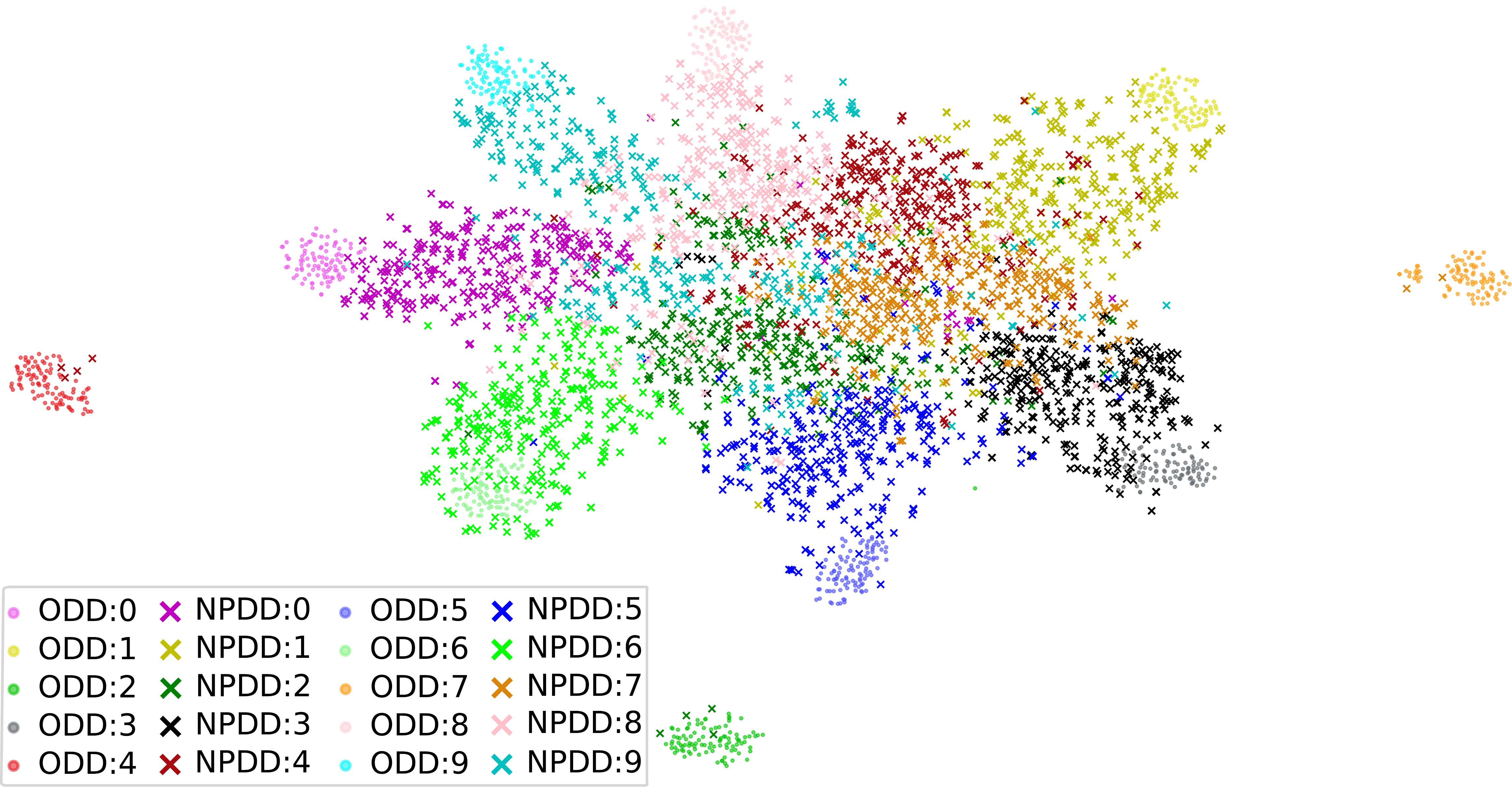}
		\caption{}
		\label{figTSNEDIG}
	\end{subfigure}
	~
	\begin{subfigure}{.35\textwidth}
%		\centering
		\includegraphics[width=.98\textwidth]{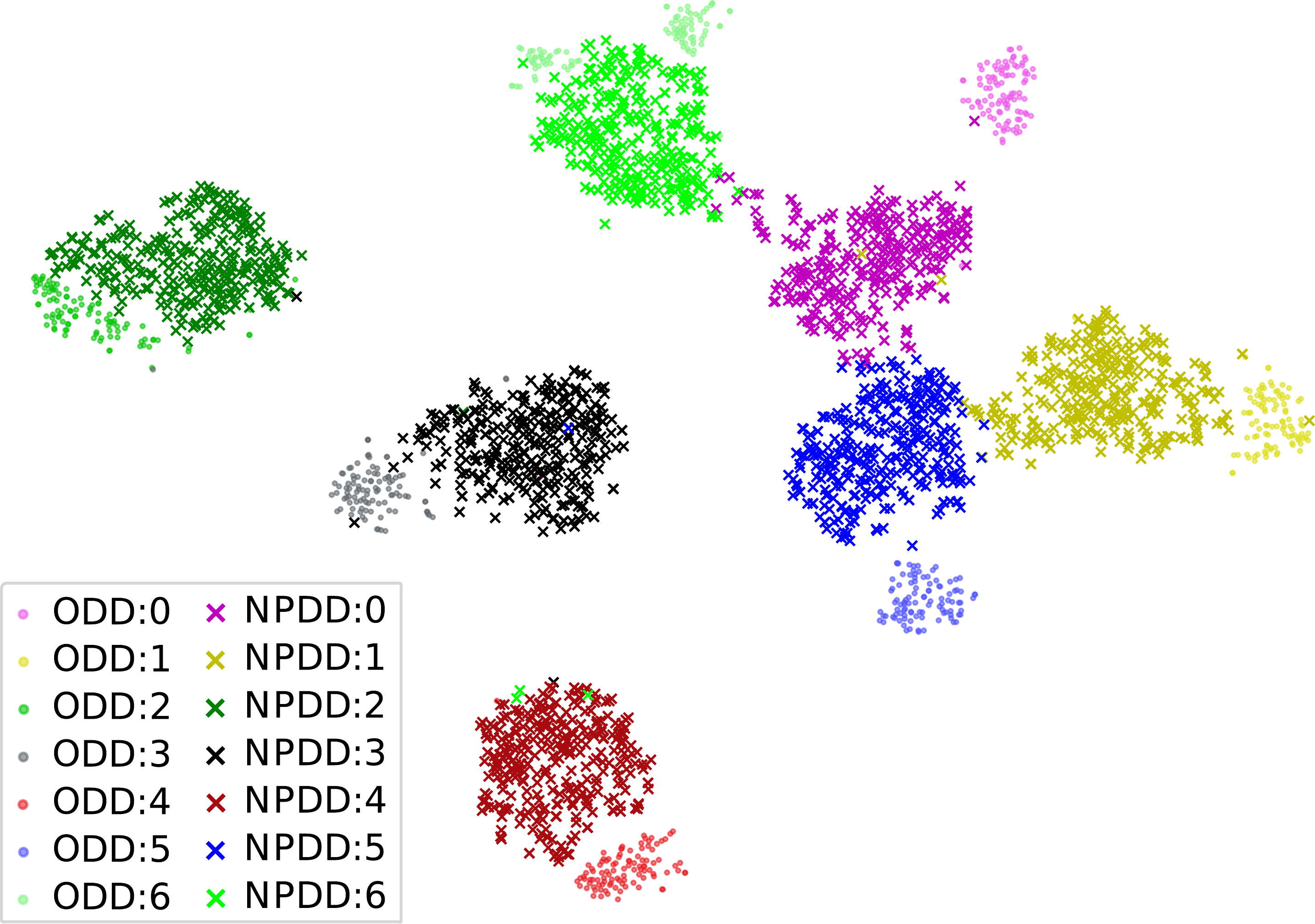}
		\caption{}
		\label{figTSNEFER}
	\end{subfigure}
	\caption{
		Mapping of the ODD-OL (dots) and NPDD-SL (crosses) to the feature space of the target network for the (a) DIG10 and (b) FER7 problems. Colors represent the different classes, but dots and crosses are represented with lighter or darker version.
	}
	\label{figTSNE}
\end{figure}

The mapping of the ODD-OL and NPDD-SL to the feature space of the target network for the handwritten digit classification and facial expression recognition problems can be visualized respectively in \autoref{figTSNEDIG} and \autoref{figTSNEFER}.

As it can be seen in Figure~\ref{figTSNE}, the classes of both problems are separated in different regions, i.e., the colors are mostly clustered. As expected, the ODD-OL images define groups that are more distinguishable and have a well-defined center since the target model was trained on these samples. Moreover, NPDD-SL images are more spread across the space, but they still cover the ODD-OL distribution fairly well. Note that due to limitations of the visualization method, the figures show only a small fraction of the NPDD-SL and ODD-OL images. Therefore, the full NPDD-SL is expected to cover a much bigger region of the space. The crosses representing NPDD-SL images are mapped to regions adjacent to dots with equivalent classes of the ODD-OL. Such results corroborate our hypothesis that NPDD-SL images can cover the necessary part in the ODD-OL input space to allow the copy attack. Therefore, one can expect that a specific classifier can have a good quality training using these natural random images with stolen labels. The NPDD-SL images that are mapped further away from the ODD-OL indicate that the copycat networks also copy the region of the space that is not of interest to the problem domain. However, this part should not affect the classification space of the problem domain if both are learned during training. 

\subsection{Copycat from the Same Architecture}
\label{secCopycatSameArch}
\begin{figure}[ht]
	\centering
	\includegraphics[width=0.9\linewidth]{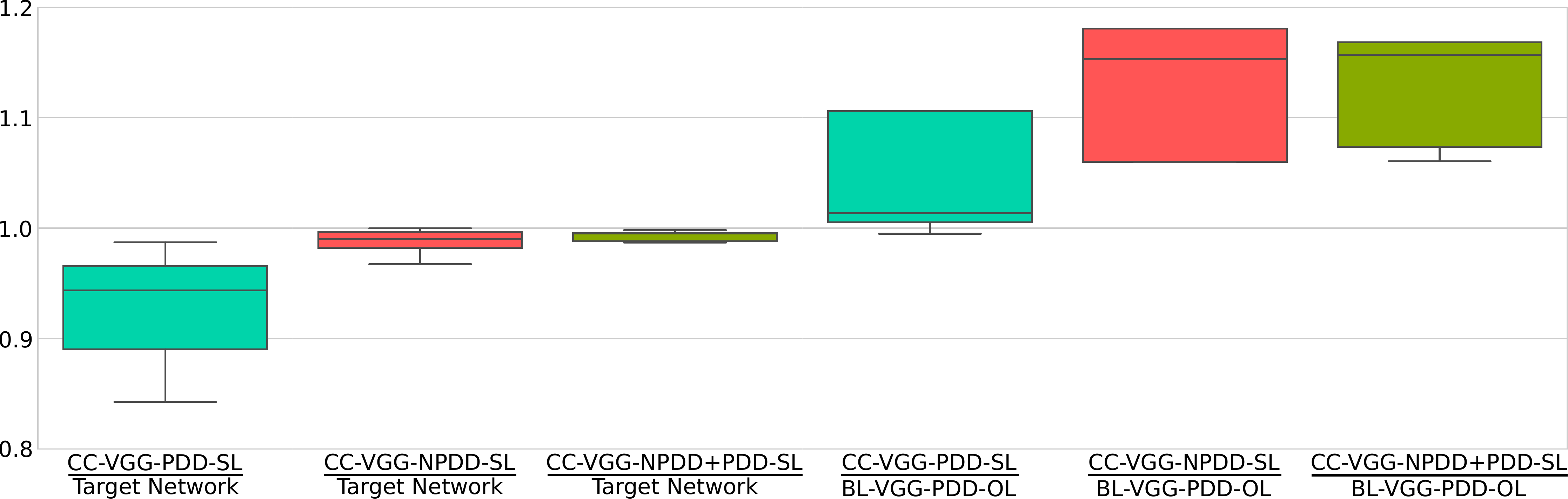}
	\caption{
		Relative accuracy of the copycats over the target network and the baseline BL-VGG-PDD-OL.
	}
	\label{fig:boxplot}
\end{figure}

The results of generating copycat networks from the same architecture as the target network (in this case, from VGG to VGG) for seven different problem domains are shown in \autoref{tabCopycatPerformance}. It shows the accuracies for the target network (BL-VGG-ODD-OL) and for the baseline (BL-VGG-PDD-OL) together with the performance of the three types of copycats
relative to the two reported accuracies. The performance relative to the target network allows measuring how close the copied model is from the original model. The performance against the baseline allows measuring the gain of the model compared to training it with a smaller dataset. 
The worst performance of the copycat with PDD-SL was on ACT101 ($58.3\%$) followed by SIG30 ($84.2\%$).
All the other copycats for these problems achieved high performances. The worst copycat with non-problem domain images was on ACT101, with performance of $96.7\%$. Comparing the CC-VGG-PDD-SL models (\autoref{tabCopycatPerformance}, 4th column) with the target network, the performances were at least $93.8\%$ on problems with 10 classes or less. The CC-VGG-NPDD-SL models (\autoref{tabCopycatPerformance}, 5th column) achieved at least $98.1\%$ of performance, except ACT101 which achieved $96.7\%$. The CC-VGG-NPDD+PDD-SL models (\autoref{tabCopycatPerformance}, 6th column) achieved at least $98.7\%$ of performance, except ACT101 which achieved $97.1\%$. As it can be seen in~\autoref{fig:boxplot}, all copycats with NPDD-SL images had an equivalent or near-equivalent performance to the target network. In addition, the copycats have equivalent or higher performance over the alternative baseline.

\begin{table}[!ht]
	\caption{Baselines accuracies (Macro Average) (TN is the Target Network, and BL is the baseline BL-VGG-PDD-OL) and copycats performance over baselines.}
	\label{tabCopycatPerformance}
	\resizebox{\textwidth}{!}{
	\begin{tabular}{c|cc|ccc|ccc}		
		\toprule
		Copycats &
		\multicolumn{2}{c|}{\textbf{Accuracies}} &
		\multicolumn{3}{c|}{\textbf{Performance over Target Network}} &
		\multicolumn{3}{c}{\centering \textbf{Performance over BL-VGG-PDD-OL}}
		\\
		CC-VGG-* & TN & BL  &
		\centering{~~~~ PDD ~~~~} & NPDD & NPDD+PDD &
		\centering{~~~~ PDD ~~~~} & NPDD & NPDD+PDD \\
		
		\midrule
		ACT101 &  $0.692$ & $0.297$ & $58.3\%$ & $96.7\%$ & $97.1\%$ & $135.9\%$ & $225.6\%$ & $226.5\%$ \\
		DIG10 & $0.996$ & $0.936$ & $94.5\%$ & $99.6\%$ & $99.7\%$ & $100.5\%$ & $106.0\%$ & $106.1\%$ \\
		FER7 & $0.887$ & $0.821$ & $93.8\%$ & $98.1\%$ & $99.4\%$ & $101.4\%$ & $106.0\%$ & $107.3\%$ \\
		GOC9 & $0.952$ & $0.821$ & $94.4\%$ & $98.4\%$ & $98.7\%$ & $110.6\%$ & $115.3\%$ & $115.7\%$ \\
		PED2 & $0.999$ & -- & $98.6\%$ & $99.7\%$ & $99.8\%$ & -- & -- & -- \\
		SHN10 & $0.924$ & -- & $98.7\%$ & $99.0\%$ & $99.1\%$ & -- & -- & -- \\
		SIG30 & $0.979$ & $0.829$ & $84.2\%$ & $100.0\%$ & $98.9\%$ & $99.5\%$ & $118.1\%$ & $116.8\%$ \\
		\midrule
		Average & -- & -- & $88.93\%$ & $98.78\%$ & $98.96\%$ & $109.58\%$ & $134.20\%$ & $134.48\%$ \\
		STD & -- & -- & $14.35\%$ & $1.14\%$ & $0.91\%$ & $15.39\%$ & $51.41\%$ & $51.66\%$ \\
		\bottomrule
	\end{tabular}
	}
\end{table}

These results show that models are vulnerable even without having images or labels from the domain of the problem. This suggests that companies should work on methods to protect their model. The attacks were able to copy models with different number of classes and from several different problem domains. The experiments showed that the copycat model has, usually, superior performance when compared to the baseline network. This suggests that it might be worthy to copy a model instead training one from scratch using a small dataset.

\begin{figure}[!ht]
	\centering
	\includegraphics[width=1.\linewidth]{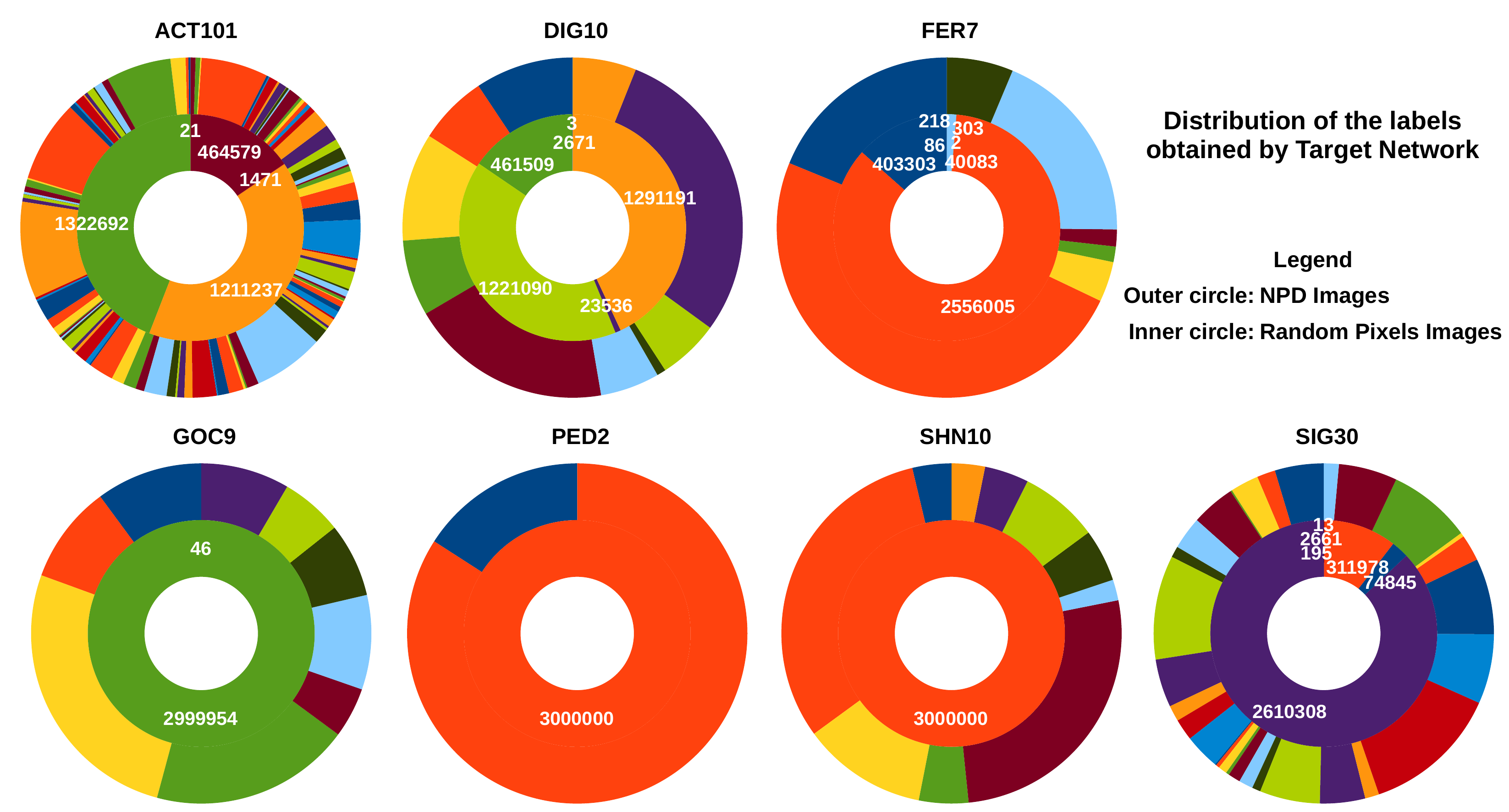}
	\caption{Distribution of labels after querying the Target Network with Non Problem Domain Images (outer circle) and Random Pixels Images (inner circle). Each pie represents a problem and each color a class (same color per class and problem).}
	\label{figImgDist}
\end{figure}

To further investigate the method, we performed an additional experiment with pixel-wise random images. Instead of using random natural images, 3 million images were artificially generated with random pixels and the same process was made with them to copy the model. After querying the model with those images, it was not possible to have a minimally balanced label distribution over all classes. As it can be seen in \autoref{figImgDist}, querying with natural random images produces a better balanced label distribution.

\subsection{Curve of Different Sizes of Training Datasets}
\label{secDataCurve}
\begin{figure}[ht]
	\centering
	\includegraphics[width=1.\linewidth]{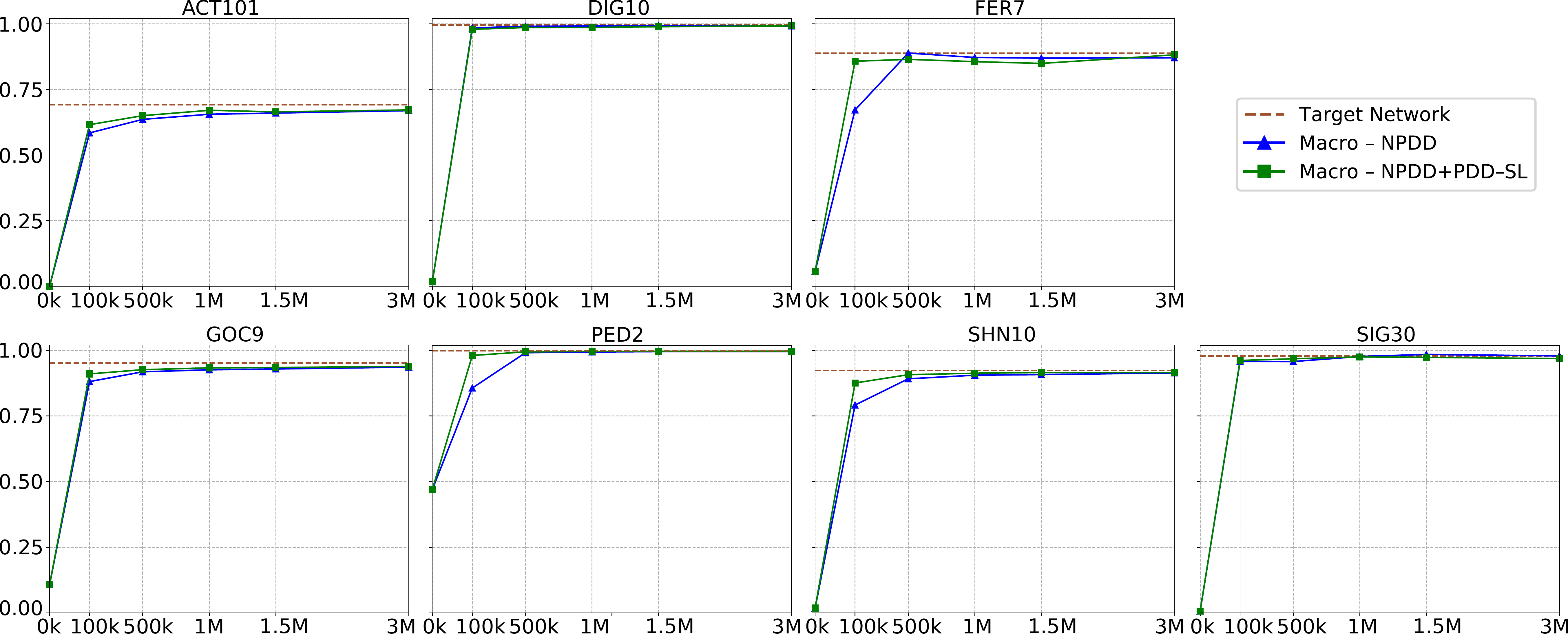}
	\caption{
		Data curve performance of copycats CC-VGG-NPDD-SL trained with 0k (network with random weights), 100k, 500k, 1M, 1.5M, 3M images of the NPDD on the seven different problems. The horizontal axis represents the fake dataset size (NPDD), and the vertical axis represents the network accuracy (Macro Average).
	}
	\label{figDataCurve}
\end{figure}

The results for the copycats CC-VGG-NPDD-SL trained with different fake dataset sizes for the seven different problems are presented in \autoref{figDataCurve}. The curves show that all copied models with any type of data and stolen labels perform much better than only initializing a model with random weights (first point of the curve, 0K). Therefore, adding samples to copy a model increases their accuracies in general. For most of the problems, a plateau can be seen in 100k images. For the ACT101, however, the plateau seems to come only after 1M images. It is worth noting that the ACT101 is a problem with much more classes than the other six problems. The  fine tuning with images from the problem domain (green curve) with stolen labels (NPDD+PDD-SL) was able to reduce the number of images to reach the plateau for some problems, when compared to the natural random images only (blue curve).

The results of this experiment indicate that the copycat models can achieve high performances with fewer images than showed in \autoref{secCopycatSameArch}. The curves showed plateaus between 100K and 1M images, which indicates that attackers might naively copy black-box models with much less queries than showed in \cite{jacson2018ijcnn}. The possibility of adding images from the problem domain to the queries seems to ensure the accuracy of the copycat models with less images. 

\subsection{Copycat from a Different Architecture}
The results of generating copycat networks from a different architecture w.r.t. the target network (from VGG to AlexNet) for seven different problem domains are shown in \autoref{tabCopycatPerformanceAlexNet}. It shows the accuracies for the target network (BL-VGG-ODD-OL) and the baseline (BL-Alex-ODD-OL) together with the performance of the copycat (CC-Alex-NPDD-SL) relative to the two reported accuracies. The performance relative to the target network (BL-VGG-ODD-OL) allows measuring how close the copied model is from the original model. However, the architecture used in the copycat might have a natural limitation due to the number of parameters of the model, which is not related to the copy itself. Therefore, the performance against the baseline (BL-Alex-ODD-OL) allows measuring how close the copycat is from a model with the same capacity. 

\begin{table}[ht]
	\caption{Performance of the copycat using AlexNet architecture.
	}
	\label{tabCopycatPerformanceAlexNet}
	\centering
	\resizebox{0.9\textwidth}{!}{%
		\begin{tabularx}{1.\textwidth}{
				>{\centering\hsize=.17\hsize}X>{\centering\hsize=.15\hsize}X
				>{\centering\hsize=.17\hsize}X>{\centering\hsize=.3\hsize}X
				>{\centering\hsize=.3\hsize}X}
			\toprule
			\multirow{2}{*}{Problem} &
			Target Network &
			AlexNet Network &
			\underline{AlexNet Copycat} \\ Target Network &
			\underline{AlexNet Copycat} \\ Alex Network
			\tabularnewline \midrule
			ACT101   &$0.692$ & $0.580$ & $78.0\%$ & $92.9\%$ \tabularnewline
			DIG10    & $0.996$ &$0.990$ & $99.2\%$ & $99.7\%$ \tabularnewline
			FER7     & $0.887$ & $0.793$ & $83.7\%$ & $93.7\%$ \tabularnewline
			GOC9      &$0.952$& $0.909$ & $91.0\%$ & $95.2\%$ \tabularnewline
			PED2      &$0.999$ & $0.908$& $97.8\%$ & $107.6\%$ \tabularnewline
			SHN10     & $0.924$& $0.897$ & $91.6\%$ & $94.4\%$ \tabularnewline
			SIG30     &$0.979$ & $0.968$& $95.8\%$ & $96.9\%$ \tabularnewline
			\bottomrule
		\end{tabularx}
	}
\end{table}

In general, the performance of the copycat model reduced when compared to the target network. The ACT101, for example, achieved a performance of only $78.0\%$, which is also the lowest performance of all problems. The FER7 problem also seems to have been affected by the limitation of the AlexNet to the problem, achieving only $83.7\%$. All other problems, presented over $91\%$ of copy. The limitation of the AlexNet model over ACT101 and FER7 is corroborated by the difference in accuracy seen when training the models with the original data (columns 2 vs. 3 of \autoref{tabCopycatPerformanceAlexNet}). When comparing the copycat models with a similar model (column 5 of \autoref{tabCopycatPerformanceAlexNet}), the copy performance is over $92.9\%$. The AlexNet's copycat on PED2 even exceeded the model trained with the ODD.

The results of this experiment show that attackers might copy models even when not knowing the architecture of the target black-box model, raising the companies concerns about protection. 
 
\subsection{Robustness of the Copycat Model}
The robustness of the model against the random factors of training was evaluated by training the same target network (BL-VGG-ODD-OL) three times on the FER7 problem and generating the corresponding three copycats (CC-VGG-NPDD-SL). The accuracies of the target networks were $90.9\%$, $90.0\%$, and $89.8\%$, 
whereas the average copy performance was $96.68\% \pm 1.32\%$. This result suggests that the different random factors of the target network training have low impact on the copy attack. Therefore, a stable performance is expected when performing a copy of a model with natural random images.

\subsection{Analysis of the Attention-Region in the Input Images}
\begin{figure}[!ht]
	\centering
	\includegraphics[width=0.6\linewidth]{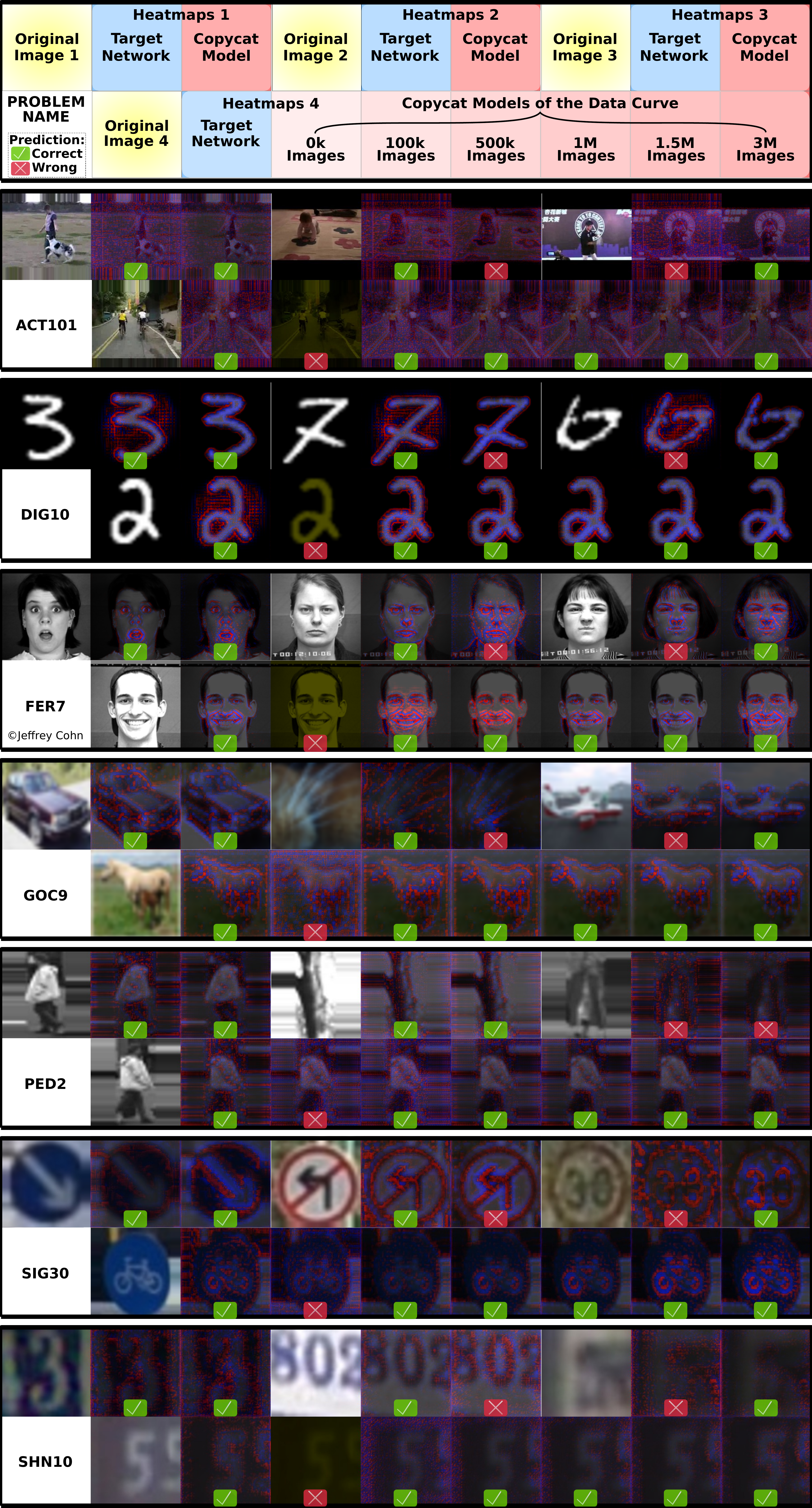}
	\caption{Heatmaps generated with LRP for three problems. The first block is an explanatory template of the subsequent blocks.}
	\label{figLRP}
\end{figure}

The heatmaps generated with LRP for the seven problems are presented in \autoref{figLRP}. Each block represents a different problem, in which the first line consists of three different predictions and the second line consists of predictions with copycat models trained with different numbers of NPDD-SL images.

In general, it can be seen that the heatmaps of the copycat models are very similar to the ones from the target network. Both models seem to focus on similar features indicating that the copycat model copied well the classification function of the target model. In the FER7 problem, for example, both models focused on the regions around the mouth and the eyes. In the DIG10, both models focused on the borders of the number and in the inner part. The most different models regarding the heatmaps are the ones from the problem ACT101. The heatmap from the target model seems to be more noisy. However, it is worth noting that the normalization process to show the colors of the heatmap might influence the results. In any case, the overall features are similar in both models. One can see, for example, the contour of the objects being highlighted in both maps. The features of interest show an evolution with the amount of images used to copy the model. In general, it can be seen that the higher the number of images used to generate the copycat, the more focused the regions of interest are in the objects of the images.

These results support our claim that the copycat models are really copying the model classification function and not just overfitting to some non relevant features of the images. The fact that both models focus on similar parts of the images shows that the copycat model has learned to mimic where to look in order to perform a correct classification. This result is more impressive when thinking that no image from the problem domain was used in the copy attack.

\subsection{Analysis of Attack Viability and APIs Costs}
\label{sec:viabilityCosts}

\begin{figure}[!ht]
	\centering
	\includegraphics[width=.7\linewidth]{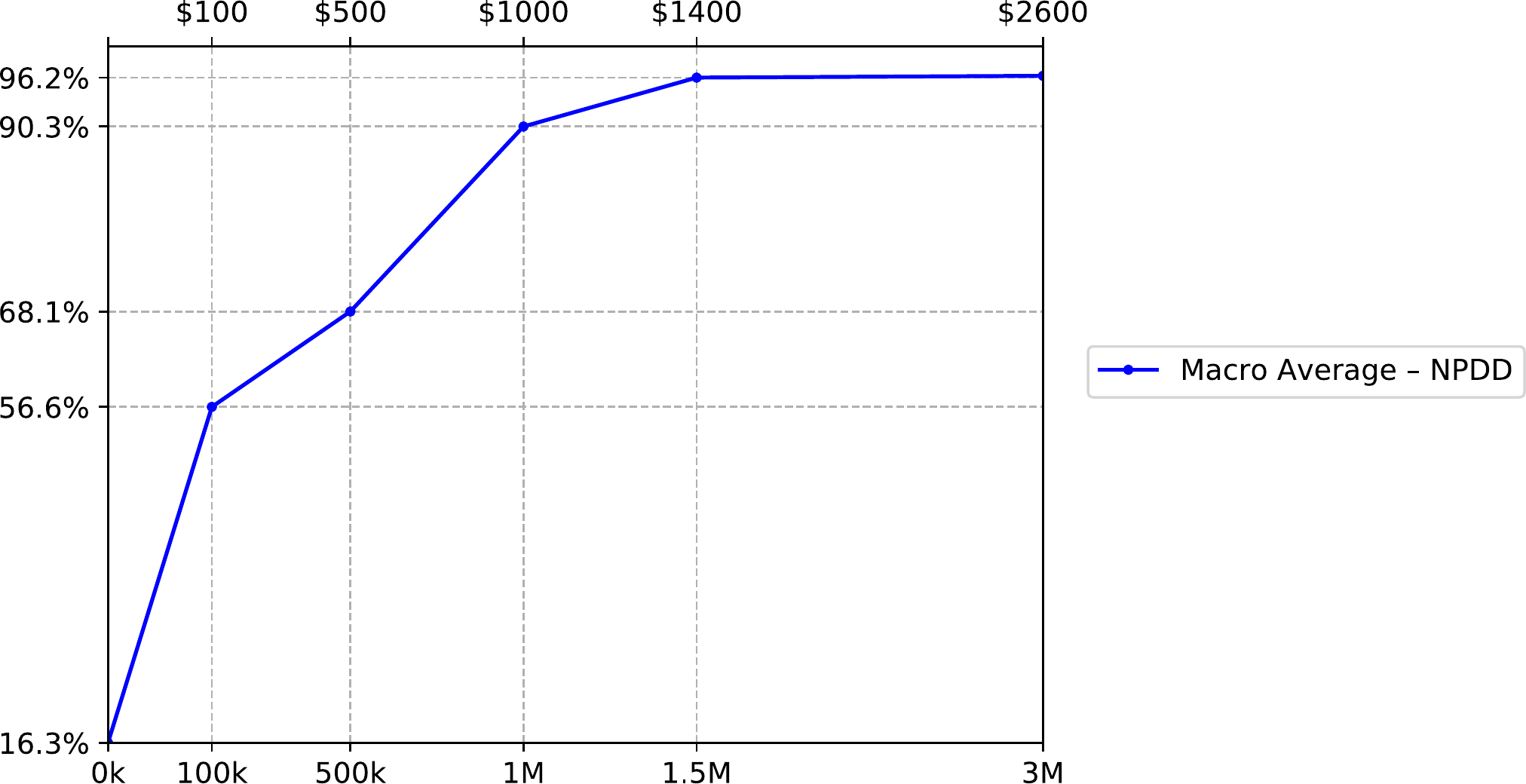}
	\caption{Performance of Copycats over Azure API on different sizes of NPDD subsets. The x-axis shows the subset size (bottom) and the pricing for generating these stolen labels (top).}
	\label{fig:AzurePerf}
\end{figure}

To assess the viability of an attack, the method is applied on a real-world API with several random subsets with different sizes of NPDD. As it can be seen in \autoref{fig:AzurePerf}, one million images were required to achieve a copy performance of over $90\%$ ($90.3\%$ with 1M, $96.2\%$ with 1.5M, and $96.4\%$ with 3M). Therefore, a realistic attack would cost at least \$1000 using only natural random images.

Given that an attack is viable when its costs are lower than acquiring and annotating data of the problem of interest, and that it is not trivial to measure the costs of generating a model of a real-world API, we measured the costs of annotating the data using the problems investigated in this work. The results are shown in \autoref{tab:labelingCosts}. As it can be seen, the price per batch (1000 images) required to get the labeling costs back shows that the current API prices are still relatively low. For instance, if the Emotion API from Microsoft Azure had costs similar to the ones of FER7 (due to the similarity of the problem), it must charge at least $\$1.90$/batch considering $1M$ of images to steal the knowledge $\left(\frac{\$1,900}{1M/1000}\right)$, but it currently charges $\$1.00$/batch for the first $1M$ transactions (it used to be $\$0.10$/batch in 2018). Given that such services might use much larger datasets to train their models, the actual costs might be even higher.

\begin{table}[!ht]
	\centering
	\caption{
		Labeling costs of ODDs and the respective minimum API's batch (1000 images) price to be charged in order to protect the model.
	}
	\label{tab:labelingCosts}
	\resizebox{.8\textwidth}{!}{%
		\begin{tabular}{ccccc}
			\toprule
			%   1          2          3               4         5
			Problems &  \#ODD  & Labeling Costs &  \#NPDD & \$ Minimum cost per batch \\
			\midrule
			ACT101   & 1782858 &    \$ 45,075   &   500k  & \$ 90.15 \\
			DIG10    &   60000 &    \$  2,000   &   100k  & \$ 20.00 \\
			FER7     &   55629 &    \$  1,900   &   500k  & \$  3.80 \\
			GOC9     &   45000 &    \$  1,575   &   100k  & \$ 15.75 \\
			PED2     &   23520 &    \$    840   &   500k  & \$  1.68 \\
			SHN10    &   47217 &    \$  1,680   &   500k  & \$  3.36 \\
			SIG30    &   31775 &    \$  1,120   &   100k  & \$ 11.20 \\
			\bottomrule
		\end{tabular}%
	}
\end{table}

\subsection{Limitations}
One of the expected limitations is the performance limit when copying from a big architecture to a smaller one. All the other copy attack methods, as well as techniques such as model distillation, suffer from this limitation too.
A limitation more specific to our proposed approach is that it needs samples with stolen labels for all classes to train a copycat model, but one cannot ensure that the target network will cover all classes of interest when labeling queries with random natural images.
In this work, however, it is shown that the proposed approach is successful in a wide range of problems, obtaining enough samples with stolen labels from all classes of interest for all the investigated problems by using natural random images.
Nonetheless, we provide evidences that it is important that the images are not completely random (fully random pixels), for which it was not possible to obtain samples with stolen labels from all classes.

%% file: sections/06-conclusion.tex
\section{Conclusion}
\label{sec:conclusion}
This work performs an extensive evaluation study aiming at providing a better understanding of the behavior of the network in black-box copy attacks using natural random images labeled by the final model output (hard labels). The study helped to consolidate a preliminary work~\cite{jacson2018ijcnn} and further explained the copycat network models. With the results presented, the community is able to better understand the vulnerabilities of their deep neural network models. 
The experimental results corroborated our hypothesis that it is possible to steal a black-box model knowledge by querying with natural random images and using only the final hard label as information.
The main findings and strengths, considering the scope of the investigated problems, can be summarized as follows.
(i) Natural random images cover well the classification space of the target networks. The classes from the original problem domain are well separated and clustered in the classification space, with samples of the natural images surrounding their clusters.
(ii) Hard labels were enough to copy the target model.
(iii) If data from the problem domain is available, the attack can be performed with fewer queries and models with better performance can be obtained.
(iv) The copy attack can be performed between different architectures. However, when copying from a big architecture to a smaller one, accuracy is limited by the number of parameters of the new architecture. 
(v) Natural random images are enough to generate a minimum balanced label distribution for the queried images. However, one cannot ensure it will always be able to sample all classes. Additionally, natural random images can generate a better balanced label distribution than using artificially generated images with random pixels.
(vi) The experiments showed copies can be performed with much less images and low costs.
(vii) Natural random images are effective to generate copycat models with similar functionalities to the black-box model. LRP showed that target and copycat models present similar activated regions.
On the other hand, Copycat has some limitations.
There is a limit to the copy performance with respect to the difference between the capacity of the black-box and the copycat architectures (known limitation from the literature).
More specific to our approach, the Copycat needs to query the API until it enough queries are labelled from all classes of interest. Therefore, although Copycat worked on several different problems using random natural images, there is no guarantee that it will work for all problems. For instance, we showed that copying with random pixels images does not work.

Overall, copycat models are effective for copying complex target models and preserving their functionalities.
Therefore, companies must invest on the investigation of methods for protecting their models, since they are more vulnerable than expected. Intuition and other academic researches might lead to the thought that one would need images from the problem domain or even access to the output probabilities to try to clone a model. However, our results showed that it is even easier to copy a model.
Future works can investigate
(i) strategies for reducing the number of queries required to perform the copy,
(ii) methods for protecting models against such type of attacks,
(iii) the performance of this attack against models trained to be robust to adversarial examples, and, finally,
(iv) the performance of the Copycat in other tasks, such as detection and segmentation.

\section*{Acknowledgements}
We thank NVIDIA Corporation for the donation of the Titan Xp used in this research.
This study was financed by the Coordena\c{c}\~ao de Aperfei\c{c}oamento de Pessoal de N\'ivel Superior - Brasil (CAPES) - Finance Code 001;
Funda\c{c}\~ao de Amparo \`a Pesquisa e Inova\c{c}\~ao do Esp\'irito Santo (FAPES, grant 594/2018), and Conselho Nacional de Desenvolvimento Cient\'ifico e Tecnol\'ogico (CNPq).